\documentclass[smallextended]{svjour3}
\usepackage{amsmath}
\smartqed
\usepackage{float}
\usepackage[T1]{fontenc}
\usepackage[latin1]{inputenc}
\usepackage{palatino}
\usepackage{color}
\usepackage{textcomp}
\usepackage{graphicx}
\usepackage[bottom =1.5cm]{geometry}
\usepackage{geometry}
\usepackage{pdfpages}
\usepackage{appendix}
\usepackage{braket}

\definecolor{checkcolor}{rgb}{0.75, 0.75, 0.75}
\newsavebox{\definitionbox}
{\end{minipage}\end{lrbox}%
\begin{center}{\colorbox{checkcolor}{\usebox{\definitionbox}}}%
\end{center}}
\title{Optimal control of traffic signals using quantum annealing}
\author{Hasham Hussain$^{\dag*}$\thanks{${^*}$H. Hussain and M. bin Javaid have contributed equally to this work} \and
        Muhammad bin Javaid$^{\dag*}$ \and Faisal Shah Khan$^\ddag$ \and Archismita Dalal$^\&$ \and Aeysha Khalique$^{\dag\S}$.
}
\institute{\dag \at School of Natural Sciences, National University of Sciences and Technology,
	H-12 Islamabad, Pakistan
	\and \ddag \at Center on Cyber Physical Systems and Department of Mathematics, Khalifa University, Abu Dhabi, UAE
	\and \& \at Institute for Quantum Science and Technology, University of Calgary, Alberta T2N 1N4, Canada \and \S \at National Centre for Physics (NCP), Shahdra Valley Road, Islamabad 44000, Pakistan
}

\date{Received: date / Accepted: date}

\date{\today}
\begin{document}

\maketitle

\begin{abstract}
Quadratic unconstrained binary optimization (QUBO) is the mathematical formalism for phrasing and solving a class of optimization problems that are combinatorial in nature. Due to their natural equivalence with the two dimensional Ising model for ferromagnetism in statistical mechanics, problems from the QUBO class can be solved on quantum annealing hardware. In this paper, we report a QUBO formatting of the problem of optimal control of time-dependent traffic signals on an artificial grid-structured road network so as to ease the flow of traffic, and the use of D-Wave Systems' quantum annealer to solve it. Since current-generation D-Wave annealers have a limited number of qubits and limited inter-qubit connectivity, we adopt a hybrid (classical/quantum) approach to this problem. As traffic flow is a continuous and evolving phenomenon, we address this time-dependent problem by adopting a workflow to generate and solve multiple problem instances periodically.
\end{abstract}
\section{Introduction}
As predicted by Moore's law \cite{Moore}, electronic computing technologies have reached the nano-meter scale and are well on their way to converging to the quantum scale within the next decade, meaning that electronic devices will have to be fabricated in a way that accounts for, and controls, quantum effects. Such devices will be instances of quantum electronics which, when assembled appropriately, will give rise to quantum circuits and hardware architectures that will constitute quantum computers. First generation quantum electronics and quantum computers are already available commercially in the form of quantum random number generators (IDquantique~\cite{IDQ}), special purpose quantum annealers (D-wave Systems~\cite{Dwave}) and limited universal quantum computers based on superconducting (Google~\cite{Google}) and ion-trap (IonQ~\cite{IonQ}) technology platforms.

The successful control of quantum features such as quantum entanglement and correlations can lead to quantum algorithms with incredible speed-ups in computational time. The famous results known as Grover's~\cite{Grover} and Shor's~\cite{Shor} algorithms, respectively, show that with a universal quantum computer, it is possible to search an unstructured database in time that is quadratic in the size of the database and that an integer can be factored in time polynomial in the number of its digits. This is in contrast to exponential time to perform each of these tasks on current 'classical' computers. Dramatic as these speed-ups are in theory, quantum computers that can successfully implement these algorithms are predicted to be at least a decade away.

However, several first generation quantum technology platforms are commercially available for early adopters to test and improve. Notable universal quantum processors include IonQ's  ion-trap based platform in which atomic qubits on a silicon chip are trapped using electromagnetic fields, and the hardware architecture connecting the qubits consists of laser pulses. Precisely tuned lasers store information on the atomic qubits, perform logical operations, including entangling the qubits. The company claims that with no fixed wires, their system can connect any two qubits with a single laser operation, increasing accuracy when compared to other technology platforms such as the one developed by Rigetti Computing. As per their Web site, Rigetti's quantum processor works in a super-cooled vacuum connected with cables driving photons that serve to calculate with qubits. Another universal quantum processor is available from IBM, the IBM Q, which comes with a user-friendly graphical user interface and a simulator for small number of qubits. The IBM Q quantum processor technology consists of superconducting transmon qubits that located in a dilution refrigerator. The focus of this paper, however, is the adiabatic quantum computing platform developed by D-wave and implemented using the quantum annealing algorithm.

Current first generation quantum annealers can be used to solve optimization problems in time that is, more often than not, less than the time it takes to solve them on classical hardware and software optimizers~\cite{McGeoch:2014:AQC:3019344,McGeoch:2013:EEA:2482767.2482797}.
A D-Wave's quantum annealer (QA) implements the following time-dependent Hamiltonian:
\begin{equation}
\label{eq:Hanneal}
H_{\rm QA}/\text{h}
= \underbrace{- A({t/t_\text{f}})\sum_i \sigma^X_i}_\text{initial Hamiltonian} 
 + \underbrace{B({t/t_\text{f}}) \left(\sum_{i} h_i \sigma^Z_i + \sum_{i>j} J_{ij} \sigma^Z_i \sigma^Z_j\right)}_\text{final Hamiltonian},
\end{equation}
where $t_\text{f}$ is the annealing time, $\sigma^X_i$ and $\sigma^Z_i$ are the Pauli matrices acting on qubit $i$, and $h_i$ and $J_{ij}$ are the qubit biases and coupling strengths, respectively (take $\hbar=1$). The optimization problem is represented in the form of the final Ising Hamiltonian~[Eq.~(\ref{eq:Hanneal})]. In the beginning of an annealing run, the system starts in the ground state of the initial Hamiltonian~[Eq.~(\ref{eq:Hanneal})], where all qubits are in a superposition of $\ket{0}$ and $\ket{1}$. During annealing the Hamiltonian of the system slowly changes from the initial to the final Hamiltonian following a preset annealing schedule given by the time-dependent functions $A(t/t_\text{f})$ and $B(t/t_\text{f})$. At the end of the anneal, the system is ideally in the ground state of the final Hamiltonian, which encodes the solution of the given optimization problem.

The QA solves the optimization problem by converting it into an Ising minimization problem, corresponding to the NP-Hard problem of maximum weighted 2-satisfiability (MAX W2SAT). MAX W2SAT can be expressed as a quadratic unconstrained binary optimization (QUBO) problem over Boolean variables, and maximization is over an objective function that can be expressed as a bilinear form over a bit string~\cite{BCM+10},
\begin{equation}\label{obj1}
{\rm Obj} := x^T\cdot Q\cdot x,
\end{equation}
where $x$ is a vector of $N$ binary variables, and $Q$ is an $N\times N$ matrix of real numbers describing the relationship between the variables. A real-valued constrained optimization problem is converted into QUBO by using penalties and binary encoding.
Many real-world problems have been formulated as QUBO problems and solved using D-Wave's QA, including optimal vehicle routing to minimize traffic congestion~\cite{NCS+17}, determining the degree of similarity between molecular structures~\cite{HA17}, and studying how transcription factor proteins bind selectively to their DNA targets~\cite{LDR+18}.

In order to run the optimization problem on D-Wave's QA, a QUBO problem is embedded into the quantum processing unit (QPU).
The architecture of the QPU is represented by a graph, with qubits as vertices and couplers as edges, following the Chimera topology~\cite{BHJ+14}. A QUBO problem is mapped onto this graph using an algorithm called minor embedding, which maps one variable to a chain of qubits~\cite{CMR14}. 
The native connectivity of the D-Wave 2000Q QA has 2048 vertices (qubits) and 6016 edges (couplers), with a working chip having fewer number due to various technical errors. In this work, we are using the system DW2000Q\_5 with 2030 active qubits and 5909 active couplers operating at temperature 13.5~mK, and a pre-defined function from D-Wave to generate embeddings.

In the following section, we briefly talk about the software package known as dwave-qbsolv, as well as the LEAP hybrid solver which is available on the cloud. The QBSolv package provides both a classical algorithm for solving QUBO instances on a PC and an all-in-one package for submitting problems to D-Wave's quantum annealer over the Internet$-$although we have only used it for the former purpose. In order to run problems on D-Wave, we have used the LEAP hybrid solver. Then we arrive at the core of the paper: defining the problem of controlling traffic signals in a road network in such a way so as to maximize traffic flow and minimize the total time wasted by cars waiting behind red signals. We explain how to approach this rather open-ended problem mathematically as a series of QUBO instances, cover the programming that was involved in this regard, and discuss the results we obtained. Using these results, we also touch upon the potential advantage, in terms of scaling, of using quantum annealing hardware to solve such a problem. All QUBO instances are solved on the D-Wave QPU DW\_2000Q available via cloud service.

\subsection{QBSolv and LEAP hybrid solver}
 A detailed description of the working of the QBSolv algorithm can be found in~\cite{BRR17}, but briefly, if we choose to solve a QUBO locally on our classical computer, we have the option of using the software package dwave-qbsolv. It implements a tabu~\cite{Glover} search algorithm for finding the minimum. Tabu search is a modified version of a local (neighborhood) search for finding the minimum of a function. It takes a potential solution and checks its immediate neighbors in search of a better solution.

Unlike other local search methods, tabu search can accept worsening moves if no improving move is available, for example, when stuck in a local minimum$-$so tabu is a less `greedy' algorithm than traditional local search. Forbidden solutions (solutions already explored) are added in a `tabu list' so as to forbid the algorithm from checking those again.



For solving problem instances on the D-Wave QPU, we have used the LEAP Hybrid solver which adopts a hybrid quantum-classical approach, suitable for large problem instances. This approach is characterized as follows. For problems too large or with arbitrary connectivity that cannot be mapped directly onto the QPU's hardware architecture, it breaks the problem into sub-problems that can be mapped into the D-Wave QPU and solved piecewise. The breaking down process is carried out classically. The LEAP solver then attempts to intelligently assign to the QPU those sub-problems which would benefit the most from running on it, while the rest are handled using the classical computing resources available on the cloud. More information about hybrid solvers may be found in~\cite{Hybrid}.

Next, we discuss our method of casting the problem of traffic signal optimization into a QUBO problem.
%
\section{Optimal control of traffic signals}
The problem at hand is: \textit{how would one control traffic signals such that maximum traffic flow is achieved in minimum time?} We made use of D-Wave's LEAP service~\cite{Dwave18} to solve certain QUBO instances of this problem. When using this service, some time is spent in latency and in the problem queue.

We point out here that while our current work is in the same spirit as the one by Neukart et al.~\cite{NCS+17}, our approach is quite different. In the work of Neukart et al., minimizing traffic congestion was achieved by assigning the most optimal route to each vehicle so as to minimize the overlapping of the routes in each vehicle's itinerary with those in other vehicles' itineraries. In our model of traffic flow, we do not consider any vehicle's route; instead, we only use information about the number of vehicles behind every signal at any given moment in time, and the speed with which they will likely travel on any given road segment. This information is then used to formulate cost functions after regular time intervals. Each cost function, when minimized, returns the most optimal configuration of traffic signals at that particular time which will cause the flow of traffic to be as smooth as possible. In short, we look to minimize traffic congestion by finding the optimal way to control the traffic signals with respect to time.

Current methods of traffic signal control include fixed cycles, coordinated control, and adaptive control. In fixed cycles, the red/green cycles for each traffic signal are fixed and do not change based on the prevailing traffic situation. In coordinated control, signal cycles are set such that drivers traveling near a certain speed will encounter green waves:  a continuous progression of green traffic signals.  Those traveling too slow or too fast will be stopped in between by red signals.  Note that coordinated signals are not the same as synchronized signals:  synchronized signals mirror each other and change at exactly the same time.

Coordinated control may be achieved in two ways:  by fixing the cycles of traffic signals based on previously collected traffic statistics or by adaptive control methods~\cite{adaptive}. These allow traffic signals to be coordinated in real time by constantly monitoring traffic patterns using cameras and/or under-pavement sensors  and  changing the cycles of each traffic signal accordingly. This lets smooth traffic flow be achieved more consistently. In our problem formulation, we aim toward achieving a kind of adaptive coordinated control.
\subsection{Problem description}
\begin{figure}[H]
\centering
\includegraphics[scale=0.07]{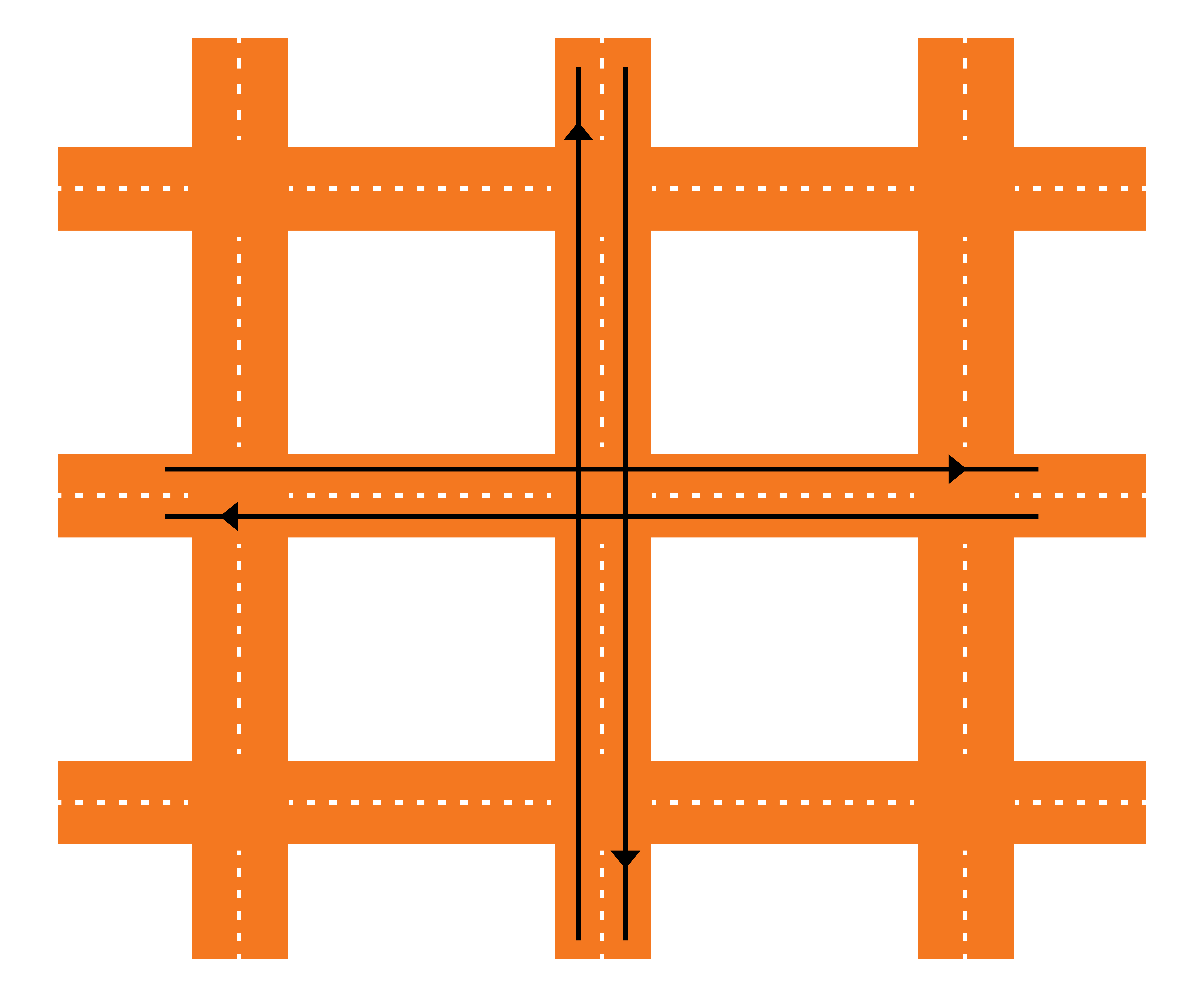}
\caption{The shape of the map}\label{graph1}
\end{figure}

\begin{figure}[H]
\centering
\includegraphics[scale=0.11]{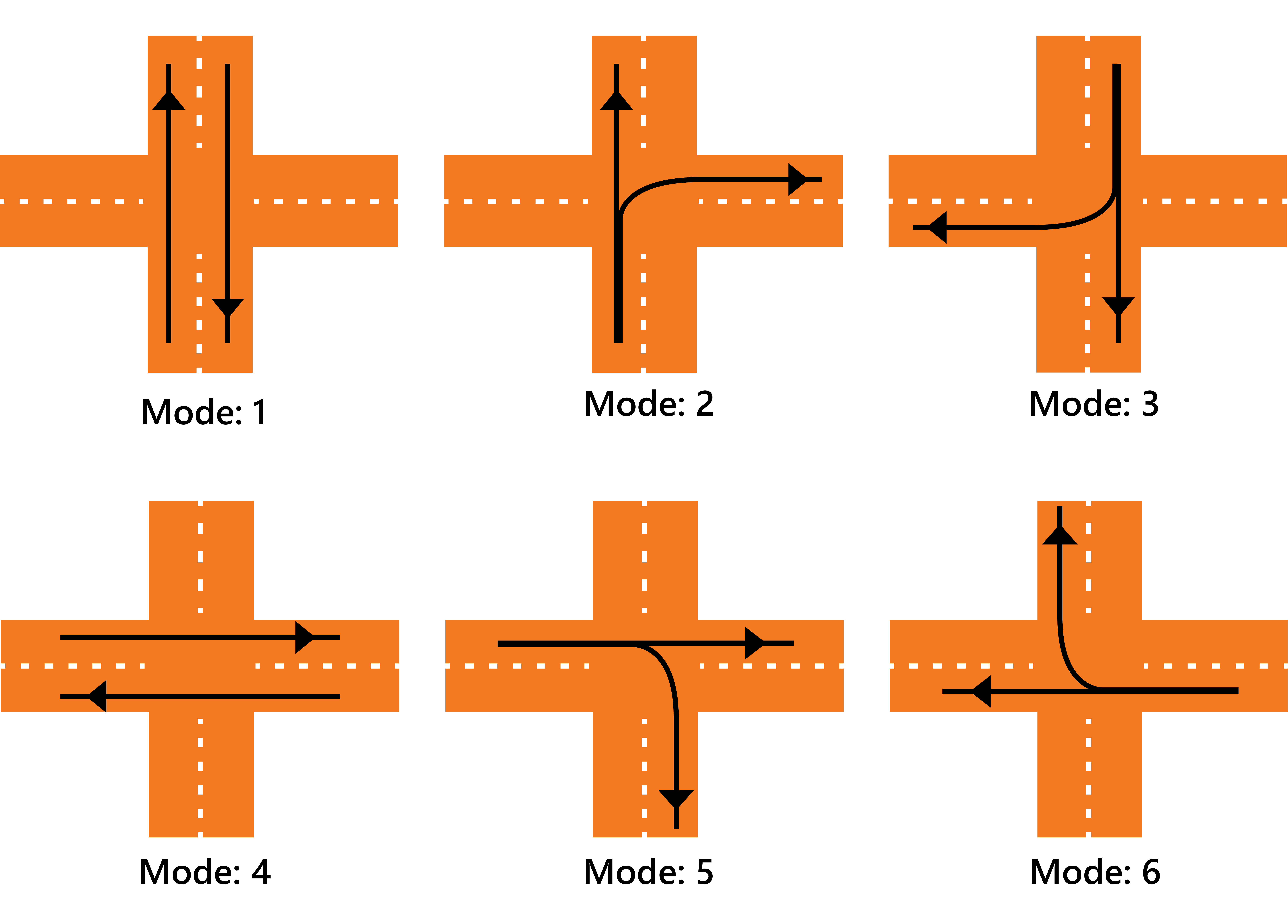}
\caption{Six possible modes of traffic flow at an intersection}\label{graph2}
\end{figure}

We assume that the road network is a square grid with identical intersections where the road segments meet; each road segment is made up of two adjacent roads with traffic moving in opposite directions, as depicted in Fig.~\ref{graph1}. As Fig.~\ref{graph2} shows, we have defined six modes in which the traffic signals may be activated at any given intersection. We consider the signal configurations in which vehicles can go straight, turn right or both. Since we consider left-hand traffic, left turns are always allowed and those that wish to turn left are not affected by the choice of traffic signal mode. The cars have the same choices at the edges and are considered only until they stay within the grid.
Let the index on the intersections (nodes) of the map runs over $i$, and the index on the modes runs over $j$. Then we have binary variables $x_{ij}$ which represent mode $j$ on node $i$, and $x_{ij}$ is 1 if mode $j$ is active on node $i$ and is equal to 0 otherwise.

Now that we have defined the binary variable to be used in the QUBO model for this problem, let us consider what we are trying to optimize in this scenario by choosing modes at each intersection so that
\begin{enumerate}
  \item the modes over the map activate in such a way so as to clear out the highest amount of stopped cars at any given time, and 
  \item the modes over the map activate in such a way so as to increase the probability that cars don't have to stop at subsequent traffic lights. The idea is that once a car has been stopped at a red signal, upon it turning green, the car should not have to also stop at the next signal.
\end{enumerate}
\textbf{Note:} There is a time-dependent aspect to this problem. Cars will keep moving and thus the locations and numbers of stopped cars will keep changing. Therefore, multiple QUBO instances will need to be formulated and solved repeatedly after fixed time intervals$-$say, $5-10$ s. Each time a QUBO instance is solved, we obtain a new configuration of modes all over the map.
\subsection{Cost function}
\begin{figure}[H]
\centering
\includegraphics[scale=0.23]{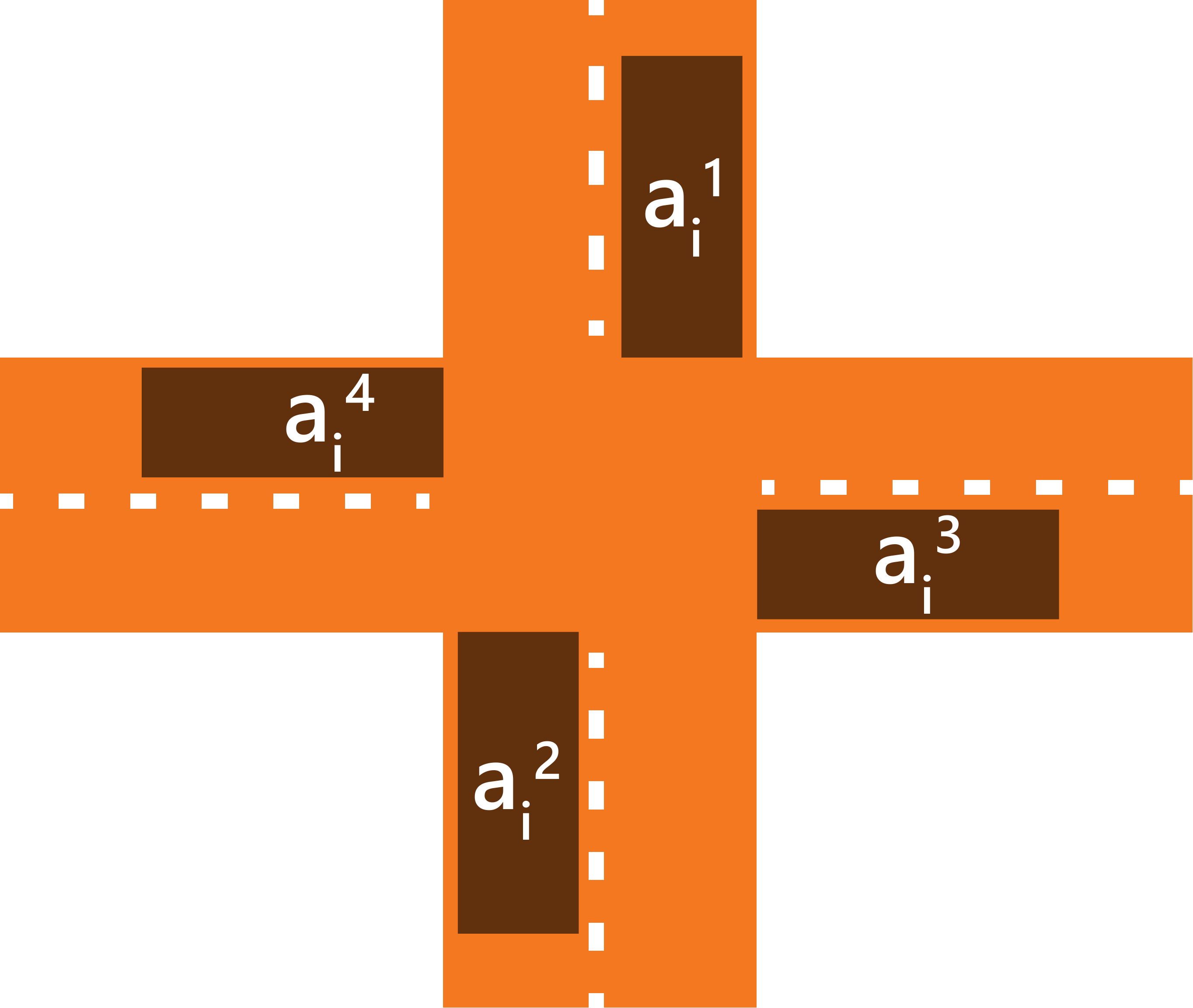}
\caption{$a_i^k$ represents the number of cars standing in the indicated position on any node $i$ (excluding those cars that wish to turn left, because they are allowed to turn left regardless of which traffic mode is active). k runs from 1 to 4, representing each incoming lane at an intersection.\label{fig3}}
\end{figure}

\begin{figure}[H]
\centering
\includegraphics[scale=0.23]{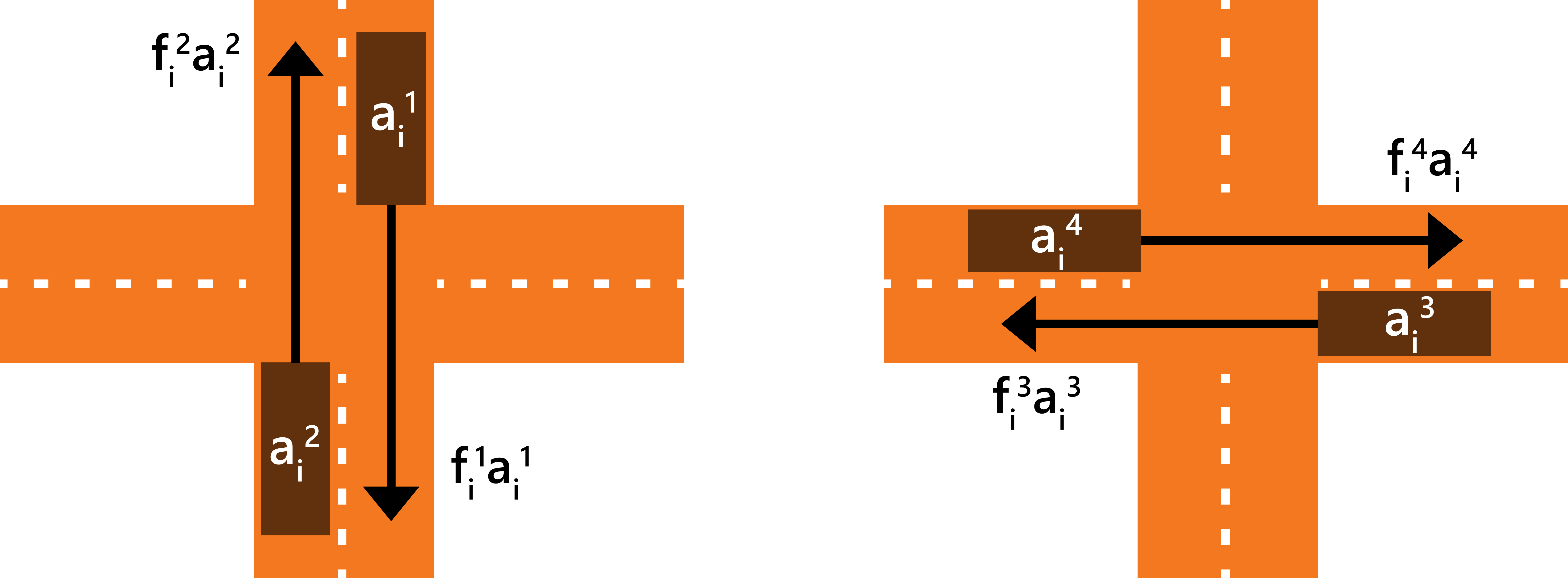}
\caption{$f_i^k$ represents the fraction of cars that want to keep traveling straight from their respective lane.}\label{fig4}
\end{figure}

To achieve goal number 1 above, we need a cost term that provides a negative cost (thus, an incentive) to activate the mode at an intersection that would clear out the most traffic. We construct the following term
\begin{equation}\label{Q1}
Q_1=-\lambda_1\sum_{i=1}^n\sum_{j=1}^6 C_{ij} x^2_{ij},
\end{equation}
where $x^2_{ij} = x_{ij}$ since it is a binary variable. $\lambda_1$ is a positive constant which we set equal to 1 for the purposes of our work. $C_{ij}$ represents the maximum number of cars that can be cleared out by activating mode $j$ on intersection $i$. For calculating $C_{ij}$, consider Fig.~\ref{fig3}.
However, not all cars will wish to go straight$-$some may want to turn right. Consider then Fig.~\ref{fig4} .
Therefore, we can calculate $C_{ij}$ as follows:
\begin{equation}
C_{i1} = f_i^1 a_i^1 + f_i^2 a_i^2
\end{equation}
\begin{equation}
C_{i2} = a_i^2
\end{equation}
\begin{equation}
C_{i3} = a_i^1
\end{equation}
\begin{equation}
C_{i4} = f_i^3 a_i^3 + f_i^4 a_i^4
\end{equation}
\begin{equation}
C_{i5} = a_i^4
\end{equation}
\begin{equation}
C_{i6} = a_i^3
\end{equation}
In the modes where cars can both turn right and go straight, then 100\% of the cars are counted and matrix elements are related to $a_i^2$. If cars were only allowed to go right then matrix elements would have been related to $(1-f_i^2)a_i^2$.

\begin{figure}[H]
\centering
\includegraphics[scale=0.05]{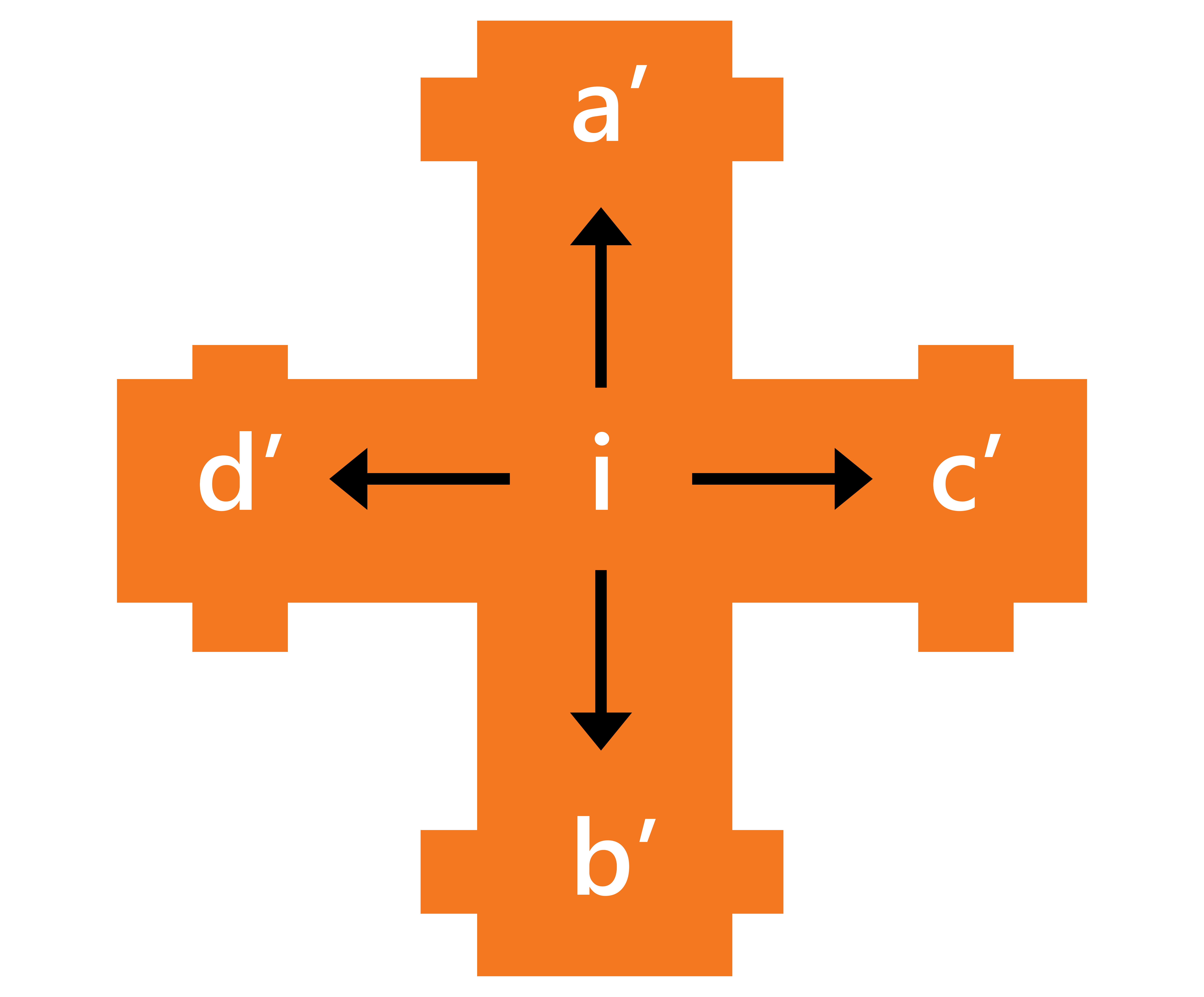}
\caption{The indices of adjacent intersections}\label{fig5}
\end{figure}

The cost of activating the mode at any intersection that will clear out the most amount of traffic will be the least, due to the negative sign in the expression. Therefore, this term fulfills its intended purpose. Of course, we will still need a constraint term that forbids all $x_{ij}$ being assigned a value of zero (no mode active at any intersection). We will introduce this constraint later. Let us first turn our attention to the term that will fulfill goal 2, namely
\begin{equation}
Q_2 = -\lambda_2 \sum_{i=1}^n\sum_{j=1}^6C_{ij} x_{ij} [\tau_{i,a'}\lambda_3C_{a'a} x_{a', a} + \tau_{i,b'}\lambda_3' C_{b'b}x_{b',b} + \tau_{i,c'}\lambda_3C_{c'c} x_{c', c} + \tau_{i,d'}\lambda_3'C_{d'd}x_{d',d}].
\end{equation}
where $i' \in \{a',b',c',d'\}$, and $i'$ refers to the intersections directly connected to the $i$ intersection. The $\tau_{ii'}$ terms are binary variables that deal with the coordination of the traffic signals on subsequent intersections $i$ and $i'$. The $C_{ij}$ terms are analogous to those in Eq.~(\ref{Q1}). The $\lambda_2$, $\lambda_3$, and $\lambda_3'$ terms are constant factors. The indices are elaborated by Fig.~\ref{fig5} and the functions of the various terms are detailed below.

Consider an intersection $i$ with mode $j$. We need to look at the corresponding intersections being influenced by this mode on intersection $i$. The objective is to find modes on the affected intersections such that a smooth flow of traffic is achieved. The intersections being influenced depend on which mode is active on intersection $i$.
\begin{figure}[H]
\centering
\includegraphics[scale=0.15]{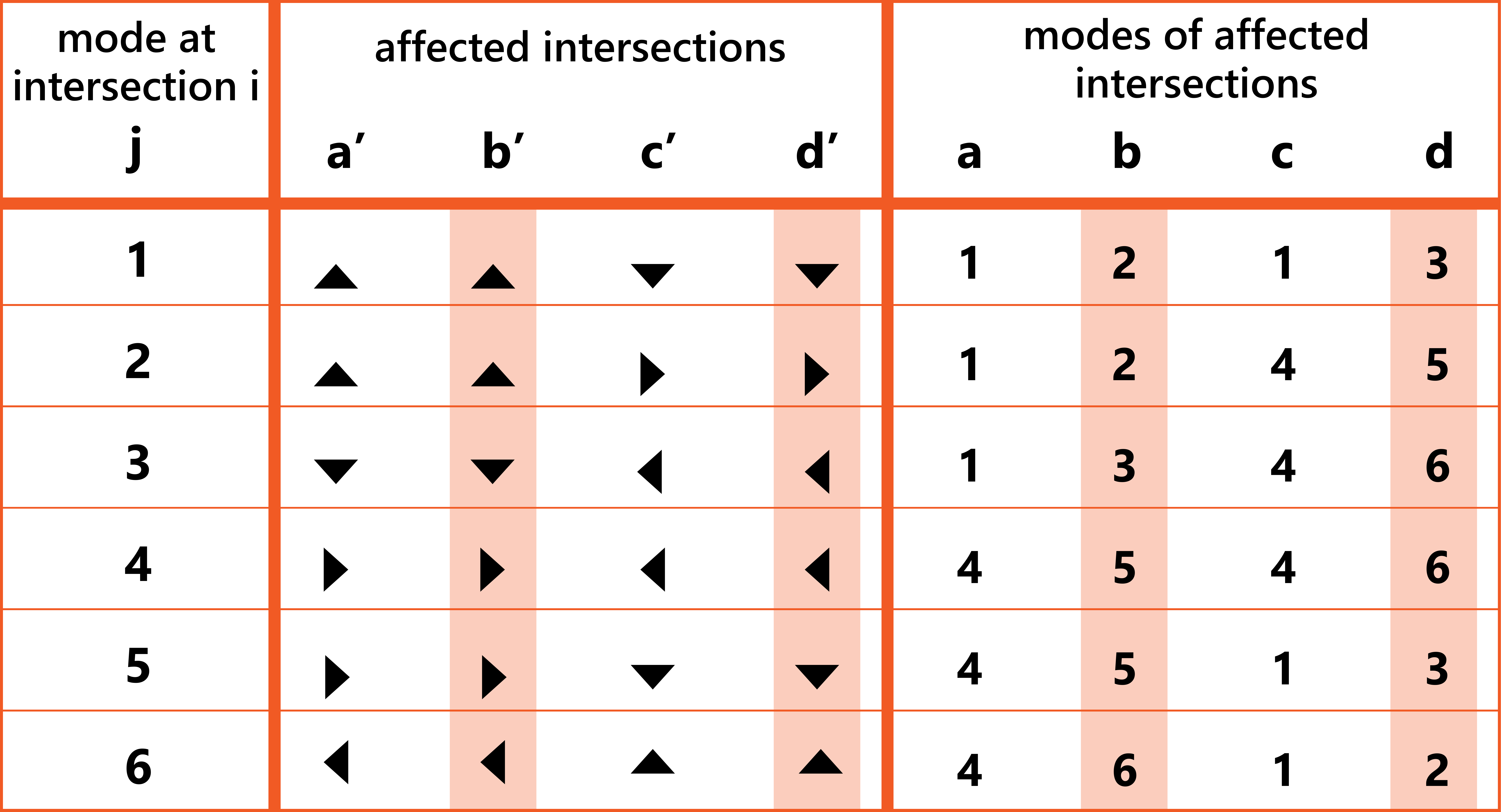}
\caption{The direction of arrows represent the intersection connected to $i$ at the indicated location (to $i$'s up, down, right, or left)}\label{fig6}
\end{figure}
$\lambda_3'$ is associated with the highlighted modes in Fig.~\ref{fig6} in which traffic is allowed to go both straight and right (namely, modes 2, 3, 5, and 6), and $\lambda_3$ is associated with the straight-going modes (modes 1 and 4). For this reason, terms representing straight-going modes, indexed by $\{a',a\}$ and $\{c',c\}$ have the same coefficient $\lambda_3$ and the other two terms representing both straight and right turning modes, indexed by $\{b',b\}$ and $\{d',d\}$ have the same coefficient $\lambda_3'$. In our simulation, the highlighted modes shown in Fig.~\ref{fig6} (at their respective intersections) have been set to receive a greater negative cost (bonus) for being selected. This is achieved by setting $\lambda_3' > \lambda_3$ (0.7 and 0.3, respectively).

In general, $\lambda_3'$ and $\lambda_3$ may be given a number of different values depending on the situation we are attempting to model. Depending on the situation, it may be more favorable to prefer the highlighted modes over the others, or vice versa$-$ and $\lambda_3'$ and $\lambda_3$ may be set accordingly. In the most general case, $\lambda_3$ and $\lambda_3'$ would be indexed over $i$ and $j$, allowing precise control over each intersection.

$\lambda_2$ represents the weight assigned to $Q_2$ in the overall cost function. In our problem we attempted traffic signal coordination on each intersection due to the uniformity of the map, and thus have kept $\lambda_2$ constant; however, in real-life situations one would value signal coordination far more for certain roads than others (e.g, long, straight highways as opposed to short side-roads), and so $\lambda_2$ would be indexed over $i$.

What we have done so far is merely synchronize the modes of nearby intersections, preferring modes in which both right-going and straight-going traffic is allowed to pass. However, there is a problem: modes synchronizing in these intersections should not be instantaneous. They should synchronize only when a car that set out from $i$ reaches the next connected intersection$-$thus achieving signal coordination leading to `green waves', a succession of green signals.

\begin{figure}[H]
\centering
\includegraphics[scale=0.3]{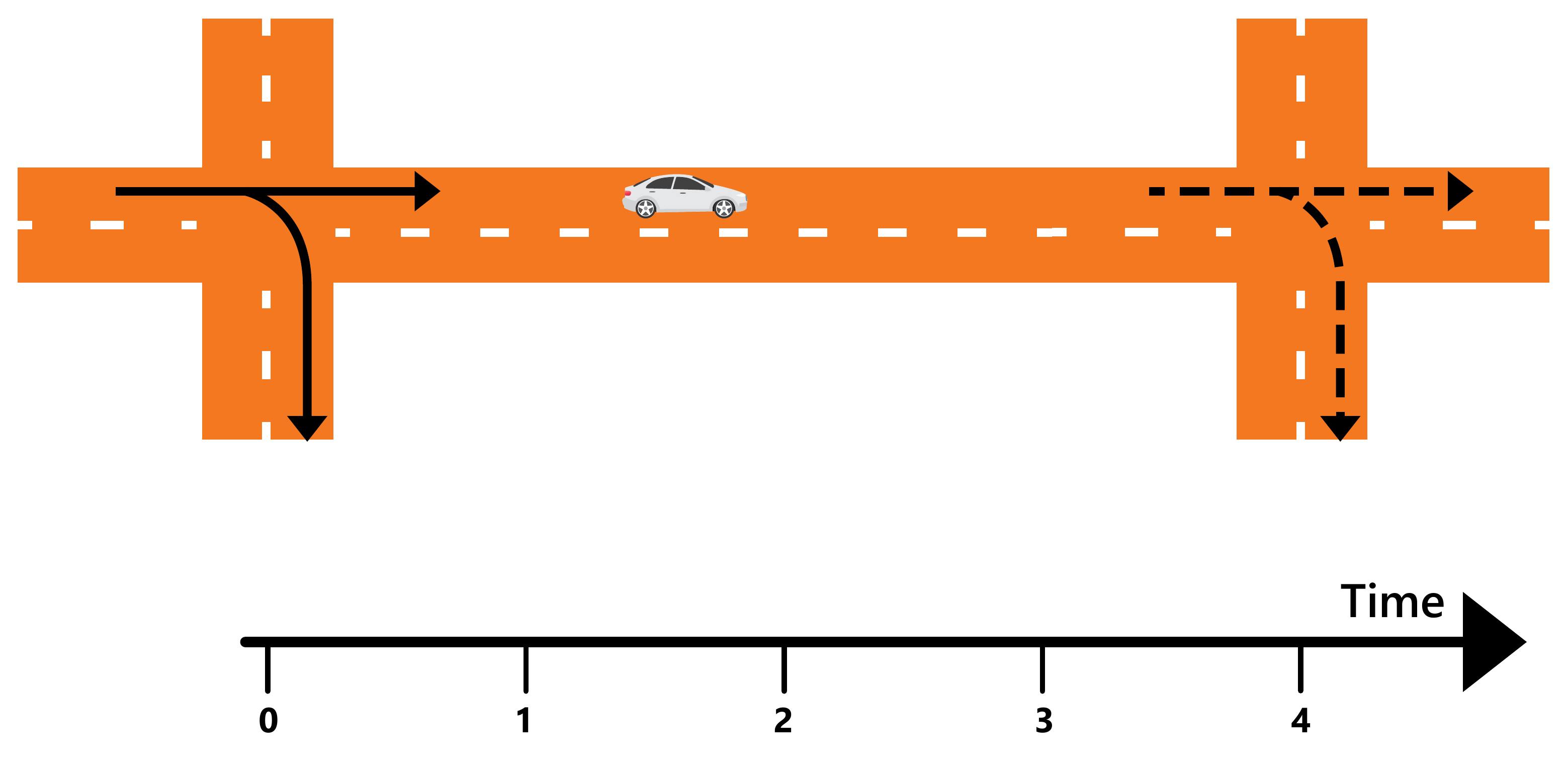}
\caption{For example: the average car takes 4 s to cross this segment; so the mode-synchronizing should occur only when the car is sufficiently close to the signal at the end of the segment}
\label{fig:sync}
\end{figure}

Recall that we are solving multiple QUBO instances repeatedly. We know the length of a road segment and the speed limit on it (assume speed limit of segment = speed of average car on that segment). Therefore, we know how long the average car will take to travel between any two intersections ($\Delta t_{ii'}$). We also know the current time $t$ that has elapsed. Using this information, we construct parameters $\tau_{ii'}$ which have value zero except when enough time has elapsed for a car to reach the next intersection under consideration, say $i'$, from an intersection $i$. In this case, $\tau_{ii'}$ is set equal to 1 for that QUBO instance. This corresponds to a bonus (negative cost, equal to $\lambda_2$) being added to the objective function if it coordinates the modes of those two intersections. Mathematically, we can represent this condition as
\begin{equation}
t\;\textrm{mod}\;\Delta t_{ii'} \approx 0
\end{equation}
where $\tau_{ii'}$ equals 1 only if this condition is satisfied. For cases where $t<\Delta t_{ii'}$, the condition
\begin{equation}
\Delta t_{ii'}-t \approx 0
\end{equation}
is used instead. These parameters $\tau_{ii'}$ ensure that mode-synchronizing only occurs when a car has had enough time to move from intersection $i$ to $i'-$otherwise, $\tau_{ii'}$ is zero and the synchronizing terms vanish (Fig~\ref{fig:sync}).
We also have $C_{ij}$ terms which contain traffic density information in this expression. As elaborated previously, a term $C_{ij}$ simply tells the number of cars that would be allowed to move if mode $j$ were active on intersection $i$. This is to ensure that if there is a conflict between mutually exclusive synchronizing options, the option that would benefit the highest number of vehicles is chosen.

Finally, we have a constraint term that ensures that one mode, and only one, is selected at each intersection, giving 
\begin{equation}
Q_3=\lambda_4 \sum_i [1 - \sum_{j=1}^6 x_{ij}]^2,
\end{equation}
where $\lambda_4$ is the penalty for violating the constraint. It should be assigned a value large enough so as to always be satisfied in the minimum of the QUBO. Notice that for any fixed $i$, the value of this term is zero only if exactly one of the $x_{ij}$ for that given $i$ is chosen to be equal to 1. Therefore, in order to avoid adding a massive penalty cost to the Objective function, the solver will be forced to choose a solution where only one mode is selected for each intersection. The full objective function for this problem, ($Q_1 + Q_2 + Q_3$), is therefore 
\begin{equation}
\label{eq:Obj}
\begin{split}
\textrm{Obj} &= -\lambda_1\sum_{i=1}^n\sum_{j=1}^6 C_{ij} x^2_{ij} -\lambda_2 \sum_{i=1}^n\sum_{j=1}^6C_{ij} x_{ij} [\tau_{i,a'}\lambda_3C_{a'a} x_{a', a} + \tau_{i,b'}\lambda_3' C_{b'b}x_{b',b} + \tau_{i,c'}\lambda_3C_{c'c} x_{c', c}\\& + \tau_{i,d'}\lambda_3'C_{d'd}x_{d',d}] + \lambda_4 \sum_i [1 - \sum_{j=1}^6 x_{ij}]^2
\end{split}
\end{equation}

$\lambda_1$, $\lambda_2$, and $\lambda_4$ are the weightages given to their respective cost function terms. Increasing $\lambda_2$, will, for example, increase the priority of mode coordination.

In the next section, we will cover the details of how we programmed this cost function. We also made a traffic simulation which used this cost function to decide the states of the traffic signals on a map. We compared the results of using this QUBO to find optimal traffic signals, with both fixed-cycle signals and signals that effectively only used the first and last terms of the Objective function (so, no signal coordination). This was to see whether there was actually a benefit to coordinating signals in this manner or not. We also compare the performance and results of the classical QBSolv tabu solver vs the hybrid quantum-classical LEAP solver which utilizes the D-Wave QPU.
\section{Programming}
We made a routine that constructs QUBO instances (Q matrices) for this problem~\cite{code}, using the following data.
\begin{itemize}
    \item[$\bullet$] Map of traffic signals. This map contains signals at intersections and the distances between connected intersections.
    \item[$\bullet$] Speed limit of each road segment (needed for calculating average time taken to travel between any two intersections)
    \item[$\bullet$] Current traffic density on the segments leading to an intersection (the $a_i^k$ values). Cars are counted as contributing to the traffic density as long as the distance between successive cars is below a certain number$-$for example, 5 m.
\end{itemize}

We now have enough information to start constructing the Q matrix from the objective function of this problem. When the program runs, it executes as follows.
\begin{itemize}
    \item[$\bullet$]  Generate a graph representing a grid map, resembling that in Fig.~\ref{fig6}. This map contains the lengths of each segment and the speed limit of each segment. We then calculate the time taken for the average car to traverse this segment using these two quantities (assume speed limit of that segment$=$speed of the average car on that segment). At the end of this process, we have a quantity for each segment that equals the average time needed to traverse it.
    \item[$\bullet$] Using the current traffic density on the road segments behind signals ($a_i^k$), compute all $C_{ij}$ values as outlined in the problem formulation.
    \item[$\bullet$] Compute the objective function associated to a Q matrix.
    \item[$\bullet$] Solve this QUBO instance using dwave-qbsolv (has the option of either solving using tabu search method or by sending to D-Wave's quantum annealer)
    \item[$\bullet$] Interpret the resulting vector of binary variables $(x_{ij})$ in terms of modes at every intersection.
\end{itemize}
Now comes the time-dependent part. We have only solved using the first term (and, of course, the constraint) of the objective function so far. What about the second term that coordinates linked intersections?

In order to incorporate that term, we constructed a simulation that places cars on the map according to the initial traffic density data information. Once a single QUBO instance has been solved according to the first and the last terms of the cost function in Eq.~(\ref{eq:Obj}), it takes the resulting modes at each intersection and moves the cars (according to the speed limit of each segment). It functions in iterations$-$and we may speed up or slow down the simulation by assigning a different value for the real-time a single iteration is supposed to represent. Once this simulation has been kick-started, we can start making use of the second term in the objective function. The simulation `knows' how much time has elapsed since its starting, and it also has the information of the time taken to traverse any segment in the map. Using this information, it keeps calculating all $\tau_{ii'}$ all over the map according to the aforementioned condition for $\tau_{ii'}$.

The simulation has the additional purpose of updating the traffic density information for the next time a QUBO instance is solved. In short, the simulation has three purposes.
\begin{itemize}
\item[$\bullet$] Calculate $\tau_{ii'}$ to be used in the objective function for the next time a QUBO instance is generated and solved.
\item[$\bullet$] Move the traffic$-$resulting in changed $C_{ij}$ for the next time a QUBO instance is generated and solved.
\item[$\bullet$] Visualize the moving traffic on the map.
\end{itemize}
It is not necessary that a QUBO instance be solved every time a single simulation iteration completes, since a single simulation iteration only represents 1~s in simulation time. We may set a QUBO instance to be solved after, for example, every 5~s in simulation time, which means after every 5 iterations. For iterations where a new QUBO is not solved, the traffic will keep on moving as per the current configuration of signal modes all over the map. When it is time to change the modes, the program will pick up the current $C_{ij}$ values from the simulation and the current $\tau_{ii'}$ and compute the Q matrix.  When this new QUBO instance is solved, the modes at each intersection will change and as a result certain cars on the map will need to stop, start moving, or keep moving as before$-$the simulation handles this.

\begin{figure}[H]
\centering
\includegraphics[scale=0.4]{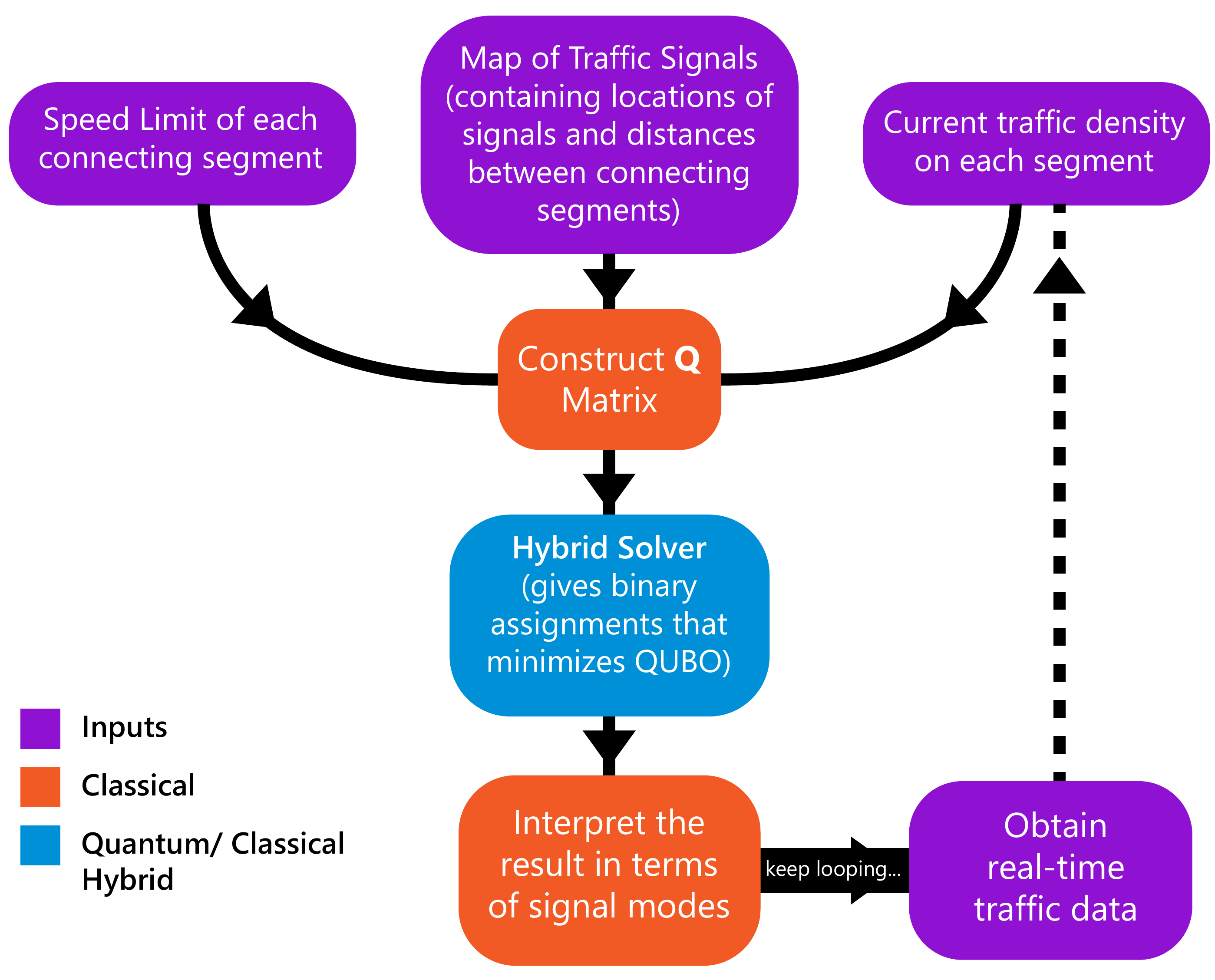}
\caption{Procedure followed by the program - since we have not tested this problem in a real-life scenario, we took real-time traffic data from the simulation after each iteration. In real life, this data would come from road cameras / sensors.}\label{fig7}
\end{figure}

The flowchart in Fig.~\ref{fig7} outlines the entire process followed by the program.
Some details about the simulations we ran are as follows. 
\begin{itemize}
    \item[$\bullet$] Number of intersections$=36$ ($6 \times 6$ grid map)
    \item[$\bullet$] Number of cars$= 4320$ (spread evenly across the map at first$-$cars keep moving / trickling out of the $6 \times 6$ grid as time passes)
    \item[$\bullet$] Possible speed limits of roads (in meters/second)$=$11, 17, 22, 28 (speed limits were randomly assigned to each road segment out of these)
    \item[$\bullet$] Length of each road segment = 1km
    \item[$\bullet$] $f_i^k = 0.7-$at every intersection, we have assumed 70 percent of incoming cars from any given direction wish to keep traveling straight, as opposed to turning right
    \item[$\bullet$] Total number of iterations run in every case = 150  (150~s, or 2.5~min, in simulation-time)
    \item[$\bullet$] $\lambda_1 = 1$, $\lambda_2 = 60$, $\lambda_3 = 0.3$, $\lambda_3' = 0.7$, $\lambda_4 = 60$. There are many possible $\lambda_2$ and $\lambda_4$ values which will also produce optimal results. As a rough guideline, both of these should be at least as large as the largest coefficient in the term $Q_1$. The first term controls traffic signals locally$-$so as to clear out the maximum number of vehicles at each intersection. The second term handles traffic signals coordination to produce green waves. If traffic signal coordination is considered more important than clearing out as many cars as possible at each intersection, then $\lambda_2$ is increased. If the opposite is true, then it is decreased. We found a value of $\lambda_2$ through trial and error so as to minimize a certain metric (total number of hours wasted). $\lambda_4$ is fixed and kept at the minimum possible value such that only one mode is selected per intersection$-$ too low a value of $\lambda_4$ results in the activation of two or more modes at the same time,  whereas too high a value results in the third term $(Q_3)$ undermining the other two and resulting in less-than-optimal signal modes.  We found this value through trial-and-error to be about 60 for this particular experiment.
    \item[$\bullet$]  A new QUBO instance was formulated and solved after every 5 iterations (5~s in simulation-time). So, the traffic signals over the map changed after every 5~s.
\end{itemize}
\subsection{Summary of the workflow}
The objective of the controlling traffic signals problem is to maximize the flow of traffic. That is, we are tasked with controlling the traffic signals in a road network over a period of time (which lights to manipulate). In order to do so, we have the following (classical) information available.
\begin{itemize}
    \item[$\bullet$] Road network map with traffic lights on the intersections.
    \item[$\bullet$] Distances between intersections and the speed limits of these segments.
    \item[$\bullet$] Number of cars stopped / moving bumper-to-bumper behind a traffic signal.
    \item[$\bullet$] The current time elapsed since the simulation began.
\end{itemize}
Now, in order to solve this problem, we divide the problem into a chronological workflow:
\begin{enumerate}
    \item Process map data, including the lengths of all segments and their respective speed limits. (Classical)
    \item Process traffic density data and information as to which signals to synchronize in this instance (if any). (Classical)
    \item Formulate the QUBO matrix. (Classical)
    \item Find a solution that provides smoothest flow of traffic across the route. (Hybrid classical/quantum if D-Wave QPU used)
    \item According to the resulting signals configuration, the simulation moves traffic on each segment resulting in updated traffic density data. The simulation also figures out which signals to coordinate for the next run. (Classical)
    \item Repeat steps $2-5$.
\end{enumerate}
Here, `Classical' refers to calculations carried out on classical machines.
\section{Results and observations}
\subsection{Quality of solutions}
\begin{figure}[H]
\centering
\includegraphics[scale=0.2]{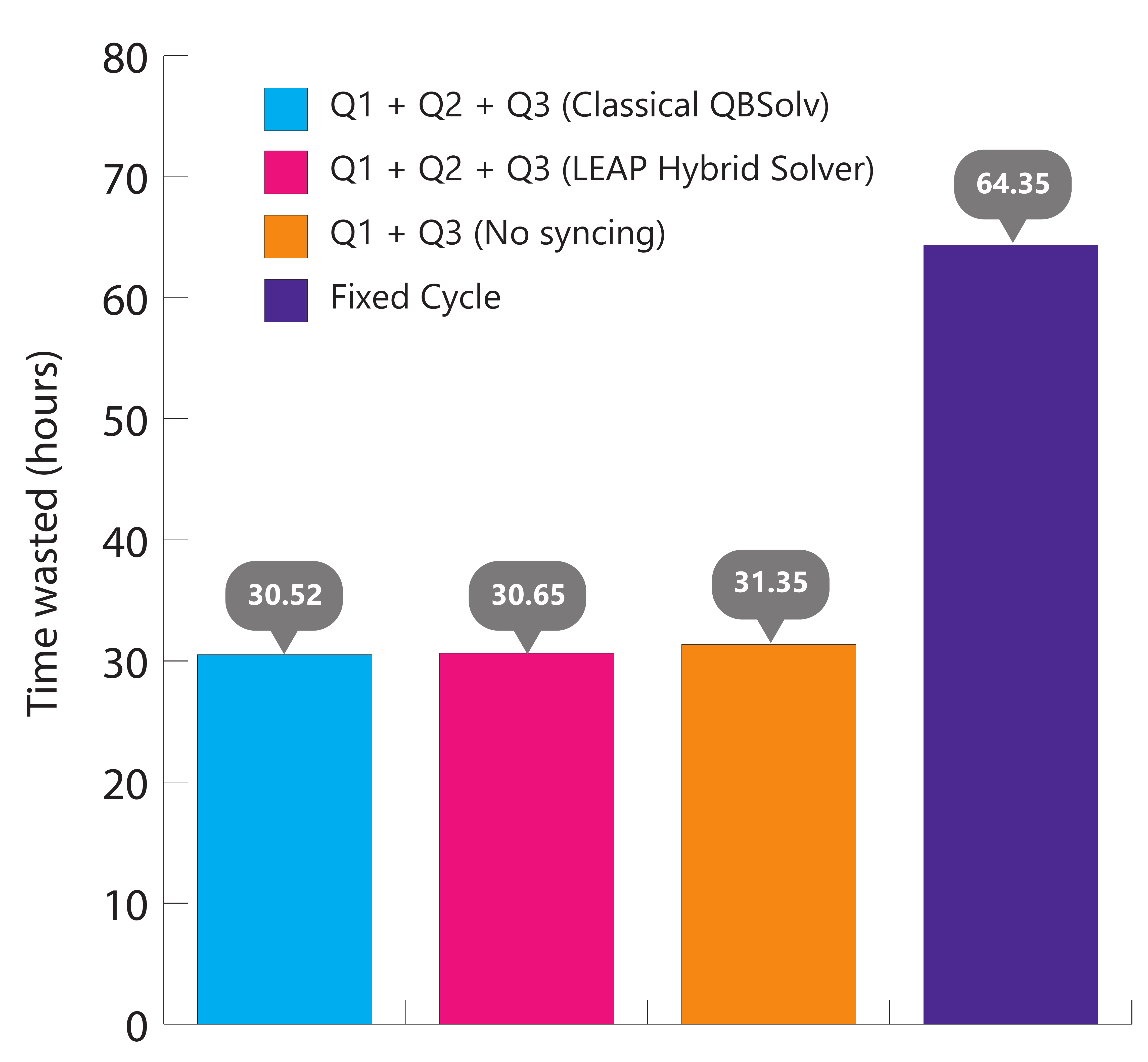}
\caption{Comparison of the metric as obtained in four different cases}\label{fig8}
\end{figure}

We require some kind of objective metric using which we can quantify the optimality of a certain method of controlling traffic signals. The metric that seems most reasonable would be the total amount of time spent waiting behind red lights by all cars, but weighted by the speed limit of the road each car will travel to after the light turns green. So, for example, a car stopped behind a highway road has its wasted time weighted more, since it could potentially travel further in the same time were it allowed to move. This formula is:
\begin{equation}
\textrm{Time Wasted} =\frac{\textrm{speed limit of next road segment}}{\textrm{maximum speed limit}}\times\textrm{N}
\end{equation}
where N is the number of cars stopped and wanting to move on to the next road segment. This metric is calculated on each intersection during every iteration, and then summed continuously as the simulation proceeds. The bar chart in Fig.~\ref{fig8} illustrates the metric obtained in four cases calculated over 150~s in simulation time each. The four cases are: full cost function solved using classical QBSolv, full cost function solved on D-Wave using LEAP hybrid solver, cost function with no coordination term $(Q_2 = 0)$, and fixed signal cycles, where signals are changed purely based on time elapsed and not on the traffic data. The latter two are run classically - as without the coordination term $Q_2$, there is no inter-dependence between the traffic signals and the problem is trivial.
The hybrid quantum-classical case was run 10 times and the average of the metric taken for those runs. The hybrid case resulted, on average, in only a marginally worse metric$-$the difference between it and the purely classical full cost-function case is 0.43\%. The slight variations are possibly due to noise in the QPU. In general, we observed that the difference in the metric between the hybrid and classical cases does not exceed 1\%. On the other hand, the classical tabu search algorithm used is completely reproducible, and produces exactly the same solutions (and thus metric) every time it is run for a certain problem with fixed parameters.

It can be observed that coordinating signals provides a benefit to commuters. In the classical QBSolv case, it saves a total of about 50~min compared to no coordination, for all 4320 cars within a simulation run-time of 150 seconds. In the hybrid quantum-classical case, the time saved is 42~min when compared to the case where signal coordination is not carried out. Coordinating signals means that more vehicles can maintain their speed without stopping while passing through a series of green waves, and thus save time overall. The fixed-cycle case resulted in the worst metric by far, since it does not adapt based on the traffic at all$-$simply cycling through the six signal modes over time on each intersection.
\subsection{Solution times}
\begin{figure}[H]
\centering
\includegraphics[scale=0.6]{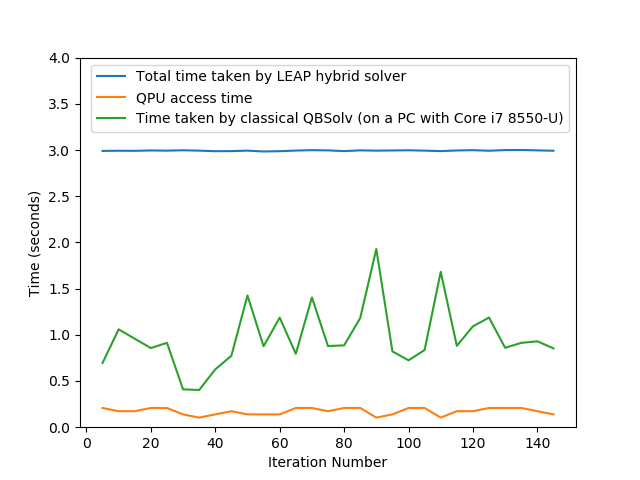}
\caption{Time taken in solving traffic signals QUBO instances on D-Wave using the LEAP hybrid solver and classical QBSolv on a PC.
\label{fig:timequboq}} 
\end{figure}

A QUBO instance was solved after every 5 iterations and the time taken to solve it measured. In the case where the problem is solved on D-Wave using the LEAP hybrid solver, the total time taken by the solver and the QPU access time are reported separately. Recall that the QPU access time is a subset of the total hybrid solver time. Figure~\ref{fig:timequboq} is a plot of the times taken for every solution instance for both the hybrid and purely classical (tabu solver) cases.
Note that there is no general large-scale upward or downward trend in the solution time as we go through all the iterations, in both the QPU and classical cases. The relatively large fluctuations in the graph for classical QBSolv may be attributed to processor scheduling.

\section{Conclusions}
We have shown that the real-world optimization problem of optimally controlling traffic signals can be formulated so as to be solvable using quantum annealing. In particular, we have shown that there is a benefit to coordinating subsequent traffic lights by including the $Q_2$ term in the cost function, so this justifies formulating the problem as a QUBO$-$as the function of the first cost function term $Q_1$ is so simple so as not to require being formulated as a combinatorial optimization problem. This benefit of signal coordination can be increased significantly if the parameters of the problem, such as the $\lambda$ values in the cost function, are fine-tuned according to the situation and only a certain subset of traffic signals are considered for coordination. Future work related to this problem involves finding an improved hybrid workflow and algorithm for solving it on D-Wave in the minimum possible time. Another point of interest is the scaling of solution time on D-Wave compared to classical solvers while varying the problem size, which in our case is determined by the size of the grid. In this work, we ran the optimization problem while keeping the problem size constant.

Naturally, the question arises as to whether it is practically feasible to solve such problems using quantum annealing$-$in particular, is there a time advantage to using quantum annealing instead of classical computers? Time is of the essence in many optimization problems$-$for example, in the traffic signals optimization problem, it is vital that we obtain an answer for a QUBO instance every $5-10$~s in the scenario that this algorithm is being used in a real-world application.

As yet, a hybrid approach to quantum annealing is outperformed in many cases by purely classical algorithms such as tabu search and simulated annealing. However, given the trend in hardware improvements, it may end up being a better choice for solving large instances of time critical problems like this one in the future. It has been suggested that the hybrid approach will remain in use for solving large problems using quantum annealing. A hybrid algorithm can leverage the strengths of each mode of computation while dealing with its weaknesses, leading to the ability to solve larger and more varied problems than would be possible for a standalone QPU~\cite{hybrid3,hybrid2}. Following are some of the current shortcomings of the D-Wave quantum annealer:
\begin{itemize}
    \item[$\bullet$] The connectivity between qubits in the physical QPU graph is limited; it is a rather sparse graph. Thus it is not possible to embed large problems directly in one go$-$as minor embeddings often do not exist for the full large problem due to the sparsity of the QPU graph. This limits the problem size that may be run directly on D-Wave, as well as the efficiency of the hybrid algorithm for solving large problems.
    \item[$\bullet$] Limited number of qubits$-$real-world problems often have several thousand variables or much more, and this combined with the aforementioned lack of connectivity between qubits leads to the same issue as in the previous point.
    \item[$\bullet$] Error rates in D-Wave's quantum annealing devices are high~\cite{SVC+18}$-$the qubits are `noisy' due to problems in manufacturing, problems in shielding, and so on. As a result, there is a large probability of qubit states collapsing prematurely due to perturbations. Due to this, a large number of resampling cycles needs to be run for each problem instance$-$adding more time required for the computation.
\end{itemize}
However, all three parameters (qubit connectivity, number of qubits, error rates) are improving with each new generation of annealers. The number of qubits have consistently gone up$-$from 128 qubits in 2012 to 2000 qubits in 2017. A 5000+ qubit version is slated for release in mid-2020, with vastly improved qubit connectivity$-$15 connections per qubit compared to 6 as in the current version~\cite{DWaveNext}.  These improvements may enhance the efficiency of D-Wave for solving optimization problems.
\begin{acknowledgements}
We acknowledge D-Wave Systems for the quantum annealing support and thank B.C. Sanders for helpful discussions. AD appreciates financial support from the Major Innovation Fund, Government of Alberta, Canada
\end{acknowledgements}
\bibliographystyle{spphys}  
\bibliography{references} 

\end{document}